\documentstyle[12pt,aasms4]{article}
\tighten

\def\lsim{\lower.5ex\hbox{$\; \buildrel < \over \sim \;$}}
\def\gsim{\lower.5ex\hbox{$\; \buildrel > \over \sim \;$}}
\def\lax    {\ifmmode{_<\atop^{\sim}}\else{${_<\atop^{\sim}}$}\fi}
\def\gax    {\ifmmode{_>\atop^{\sim}}\else{${_>\atop^{\sim}}$}\fi}

\def\gtorder{\mathrel{\raise.3ex\hbox{$>$}\mkern-14mu
             \lower0.6ex\hbox{$\sim$}}}
\def\ltorder{\mathrel{\raise.3ex\hbox{$<$}\mkern-14mu
             \lower0.6ex\hbox{$\sim$}}}

\def\pmb#1{\setbox0=\hbox{#1}%
  \kern-0.015em\copy0\kern-\wd0
  \kern0.03em\copy0\kern-\wd0
  \kern-0.015em\raise0.0433em\box0 }

\begin{document}

\title{ The Global-Normal Disk Oscillations and 
the Persistent Low Frequency QPO in X-ray Binaries}

\author{Lev Titarchuk}
\affil{NASA/ Goddard Space
Flight Center, Greenbelt MD 20771;  Center for Earth Observing and Space
Research in George Mason University; Naval Research Laboratory; USA;
lev@lheapop.gsfc.nasa.gov; lev@xip.nrl.navy.mil}

\author{Vladimir Osherovich}
\affil{NASA/Goddard Space Flight Center/RITSS, Greenbelt MD 20771 USA;
vladimir@urap.gsfc.nasa.gov}

\vskip 0.5 truecm


\begin{abstract}
We suggest that  persistent low-frequency quasi-periodic oscillations (QPOs)
detected in 
 the black hole (BH) sources  XTE J1118+480, GRO J1655-40 LMC X-1 
 at $\sim 0.1$ Hz, and  QPOs in HZ Her/Her X-1 at $\sim 0.05$ Hz  and
in Neutron Star (NS) binaries 4U 1323-62, 4U 1746-31 and 
EXO 0748-76 at $\sim 1$ Hz are caused by the global disk oscillations in the
direction normal to the disk (normal mode).
We argue that these  disk oscillations are a result of the gravitational
interaction  between the central compact object and the disk. A small displacement of the
disk from the equatorial  plane results  in a linear gravitational restoring
force  opposite to this displacement.  Our analysis shows that 
the frequency of this mode is a function of  the mass of the central 
object and it also depends on the inner and outer radii of the disk
which in turn are related to the rotation period of the binary system.
We derive an analytical formula for the frequency of the normal disk mode
and show that these frequencies can be related to 
the persistent lower QPO frequencies observed in the NS and BH  sources.
We offer a new independent approach to the black hole 
mass determination by interpreting this low QPO frequency as the global disk
oscillation frequency. The implementation of this method combined with  the  independent 
method  which uses the X-ray energy spectra (Shrader \& Titarchuk 1999)
results in stringent constraints for the black hole masses.  

\end{abstract}


\section{Introduction}

In  previous papers (Osherovich \& Titarchuk 1999;  
Titarchuk, Osherovich 1999, hereafter TO99; 
Titarchuk, Osherovich \& Kuznetsov 1999), 
 we have suggested a new Two-oscillator (TO)  model which have lead 
to the classification of quasiperiodic oscillations (QPO) in frequency range 
from $\sim 1$ Hz to $\sim$ kilohertz. 
All oscillations in the TO model are related to the local 
conditions in the disk. 

The classification offered by the TO model does not include the  persistent 
low frequency oscillations ($\sim 10^{-2}- 1$ Hz) which have been  observed in
many neutron star and 
black hole candidates systems (see reviews by van der Klis 1995 and 2000).
We suggest in this {\it Letter} that these persistent low frequency oscillations
are related to the global disk oscillation under the influence of the 
gravitational force of the central object -- in contrast to the variable
frequencies which are characteristic of the local properties of 
the disk and the magnetosphere.  Below we show that these persistent
frequencies carry information 
about the system as a whole, namely,  information about the size of a disk 
and a mass of the central object.  

Persistent low frequency quasi-periodic oscillation (QPO)  with $\sim 1$ Hz
frequencies were discovered in neutron star binaries: 4U 1323-62 
by Jonker, van der Klis \& Wijnands (1999), EXO 0748-676 by
Homan et al. (1999, hereafter H99) and  4U 1746-31 by Jonker et al. (2000).
Recently Boroson et al. (2000) have found a significant QPO feature with 
$\sim 0.05$ Hz in the power density spectrum (PDS) of UV flux from HZ Her/Her
X-1 system.  Similar persistent QPO features with $\sim 0.1$ Hz frequencies
 have been  found in the transient source XTE J1118+480 in the far-ultraviolet
 and  optical (Haswell et al. 2000) wavelengths.
The QPO feature with $\sim 0.08$ Hz in XTE J1118+480 has been also  detected 
in the Proportional Counter Array on Rossi X-ray Timing Explorer (PCA/RXTE)
data (Revnivtsev, Sunyaev \& Borozdin 2000; Wood et al. 2000).
 Ebisawa, Mitsuda, \& Inoue (1989) have also found a narrow 0.08 Hz QPO peak 
in the LMC X-1 power density spectrum  in the 1-16 keV range.

It has been emphasized by H99 that several properties of the
low frequency QPO,  particularly 
its frequency and its relatively unchanged persistence during bursts and dips
are remarkably similar among different stars.
We suggest  these QPOs are caused by vertical oscillations of disk
which we call  the Global Disk Mode (GDM) oscillations.
%

  In section 2 we  derive of the GDM frequency. We compare the calculations of the GDM  with the QPO observations
in BH sources and present a new method of the black hole mass determination
using the derived GDM oscillation frequency in section 3.   
Comparisons with the observations in NS sources are made in section 4.
Conclusions follow in section 5.

\section{Global Disk Mode Oscillation. Formulation of the Problem and Solution}

We consider the oscillations of the disk as the whole body under the influence
of  gravity of the central compact object shown as a black circle in Figure 1.
We approximate the surface density distribution in the geometrically thin
Shakura-Sunyaev
disk (Shakura \& Sunyaev 1973) by the formula 
\begin{equation}
\Sigma=\Sigma_0={\rm constant}~~~~~~~~~~ {\rm for}~~~~~R_{\rm in}\leq R\leq
R_{\rm adj}
\end{equation}
and   
\begin{equation}
\Sigma=\Sigma_0\left({R\over{R_{\rm adj}}}\right)^{-\gamma}~~~~~~~~~~ 
{\rm for}~~~~~R_{\rm adj}\leq R\leq R_{\rm out},
\end{equation}
where $R_{\rm in}$ is the innermost radius of the disk, $R_{\rm adj}$ is 
an adjustment radius in the disk and $R_{\rm out}$ is the outer radius of the
disk.  
Titarchuk, Lapidus \& Muslimov (1998; see also Titarchuk \& Osherovich 2000, 
hereafter TO00)
calculated the size of the transition disk region between the Keplerian disk and
the inner disk edge and they found that the typical radius of the 
adjustment radius depending on the effective Reynold's number is within
$(2-3)R_{\rm in}$.

 We assume that the disk as a whole is perturbed and it deviates from
equatorial plane by a small distance $h$. 
In Figure 1 we have enlarged that distance for the purpose of illustration.
The restoring force $F(h)$
caused by gravitational attraction of the central object is 
\begin{equation}
F(h)=GM_x\pi h\int_{R_{\rm in}}^{R_{\rm out}}
{{\Sigma 2RdR}\over{(h^2+R^2)^{3/2}}},
\end{equation}
where $M_x$ is the mass of the central object and $G$ is the Gravitational
constant.
After integration using the density distribution (1-2) with assumptions 
that $R_{\rm in},~R_{\rm adj}\ll R_{\rm out}$ and $h\ll R_{\rm in}$ we get
\begin{equation}
F(h)\approx{{2\pi GM_x\Sigma_0 h}\over{R_{\rm in}}}
\left[1-{{\gamma}\over{(\gamma+1)}}{{R_{\rm in}}\over{R_{\rm adj}}}\right].
\end{equation} 
The mass of the disk $M_d$ from $R_{\rm in}$ to $R_{\rm out}$ is
\begin{equation}
M_d=2\pi\int_{R_{\rm in}}^{R_{\rm out}}\Sigma RdR
\approx {{2\pi \Sigma_0 R_{\rm adj}^2}\over{(2-\gamma)}} 
\left({{R_{\rm out}}\over{R_{\rm adj}}}\right)^{2-\gamma}~~~~
{\rm for~\gamma\ne 2}.
\end{equation}

The  vertical oscillations of the disk as a whole  can be described 
by  the following equation of motion
\begin{equation}
M_d\ddot{h}~+~F=0,
\end{equation} 
which can be written in the form 
\begin{equation}
\ddot {h}~+~\omega_0^2h=0,~~~~{\rm and}~~~~
\omega_0^2=(2\pi\times\nu_0)^2={{F}\over{h M_d}}.
\end{equation}
%

Using Eqs. (4-5, 7) we present  the GMD frequency as 
\begin{equation}
\nu_0={{2.2\times 10^3~{\rm Hz}}\over{m}}
\left\{{{(2-\gamma)}\over{x_{\rm in}^3}}
{{[1-\gamma/(\gamma+1)r_{\rm adj}]}\over{r_{\rm adj}^{\gamma}
r_{\rm out}^{2-\gamma}}}\right\}^{1/2}, 
\end{equation}
where $x_{\rm in}=R_{\rm in}/3R_S$,  $r_{\rm out}=R_{\rm out}/R_{\rm in}$,
$r_{\rm adj}=R_{\rm adj}/R_{in}$, $m=M_x/M_{\odot}$ and $R_S=2GM_x/c^2$ is 
the Schwarzschild radius.

According to the Shakura-Sunyaev disk model (1973) the index $\gamma$ of the
surface density  can be either $3/5$ or $3/4$.   For $\gamma=3/5$, and  for
typical parameter values of the disk model around neutron stars
$m=M/M_{\odot}=1.4$, $x_{\rm in}=1$, $r_{\rm adj}=3$ and $r_{\rm out}=10^4$
formula (8) reads $\nu_0\approx 2$ Hz. For $\gamma=3/4$ the value of 
$\nu_0$ is also close to 2 Hz.
 
The size of the disk $R_{\rm out}$ can be estimated using the size of the Roche
lobe $R_L$. Following De Jong, Paradijs \& Augusteijn (1996), we assume that
the relative size of the accretion disk f=$R_{\rm out}/R_L$ is approximately
0.5. De Jong et al. adopted this empirical size from eclipse-mapping
observations on cataclysmic variables (Rutten et al. 1992).
Having in mind that according to Paczynski (1967) 
\begin{equation}
R_L=0.46a\left({{M_x}\over{M_x+M_{\rm opt}}}\right)^{1/3}
\end{equation} 
and using the third Kepler law relating the distance $a$ between the 
optical and X-ray counterparts of the binary, the masses ($M_x$ and 
$M_{\rm opt}$) and the rotational period of the binary $P$
\begin{equation}
a= \left({P\over{2\pi}}\right)^{2/3}[G(M_x+M_{opt})]^{1/3}
\end{equation} 
we find 
\begin{equation}
R_{\rm out}\approx 0.5R_L=0.23(GM_x)^{1/3}\left({P\over{2\pi}}\right)^{2/3}.
\end{equation}
For periods of the binary $P_3$ measured in units of three hours and 
masses of the X-ray source $m_x$ measured in the solar units, we 
have the following size of the disk
\begin{equation}
R_{\rm out}=1.7\times 10^{10}m^{1/3}P_3^{2/3}~{\rm cm}
\end{equation}
and
\begin{equation}
r_{\rm out}=1.92\times 10^{4}m^{-2/3}P_3^{2/3}x_{\rm in}^{-1}.
\end{equation}

But in fact, fitting the UV data obtained from IUE, Howarth \& Wilson (1983
hereafter HW83) 
found that  the  size of the disk in  HZ Her/Her X-1 is   approximately 
1.5 times larger than that  given by formula (12).  
 They gave $R_{\rm out}=(1.69\pm 0.25)\times 10^{11}$ cm  instead of 
$R_{\rm out}=1.08\times 10^{11}$ cm obtained from formula (12) 
for P=1.7 d and m=1.4. Thus an uncertainty of a factor 1.5 exists in the
estimate of the disk size by formula (12). 

Using (13) for the value of $r_{\rm out}$ and $\gamma=3/5$, we get 
the GDM frequency  $\nu_0$ from (8) as  
\begin{equation}
\nu_0\approx 2~{\rm Hz}~x_{\rm in}^{-8/15}m^{-8/15}P_3^{-7/15}
r_{\rm adj}^{-0.3}. 
\end{equation}

\section{ Black Hole Mass  Determination and  the Global Mode Oscillation
Frequency}

Given the success of the Bulk Motion Comptonization model 
[Titarchuk, Mastichiadis \& Kylafis 1997;
Titarchuk \& Zannias 1998;  Shrader \& Titarchuk 1998; Laurent \& Titarchuk
1999; Borozdin et al. 1999; hereafter  BOR99)
in reproducing the high-soft state continuum,
Shrader \& Titarchuk (1999, hereafter ShT99) have investigated the potential
predictive power of this model. From the spectral shape
and normalization, one can calculate an effective disk radius and the
mass-to-distance  ratio $m/d$. Other quantities  such as  a black hole mass and
mass-accretion rate can be determined if the distance or mass are known
independently.  This is dependent upon an additional factor, $T_h$,
the so called ``hardening factor''  which represents the ratio of
color-to-effective temperature. If one can identify several ``calibrators'',
i.e., sources for which distance and mass are
well determined,  the model can be applied  to  derive $T_h$ over the available
dynamic range in luminosity. Three such sources are 1) GRO~J1655-40, for which
 extensive analysis was presented in  previous papers (BOR99 and ShT99), 
2) LMC~X-1 for which the distance is well constrained, and 3) Nova Muscae 1991,
 for which the binary  parameters are reasonably constrained.
In the latter two cases,  the hardening factor, $T_h\simeq2.6$, was used 
which was previously derived from  analysis of GRO~J1655-40 (BOR99). 
The temperature-flux curve presented in BOR99 and ShT99
provide  a self-consistency test of this assumption.  

Thus, using the energy spectra of the high-soft state only, 
one can get strong constraints on the BH masses. Here 
we have verified this energy spectrum method using the global
oscillation frequency model for LMC X-1 and GRO J1655 sources and
then we determine the BH mass for XTE J1118+480 using the persistent QPO
frequency $\sim 0.1$ Hz discovered by independent groups in the different
energy bands.     

\centerline{\it LMC X-1}

Shrader \& Titarchuk (1999) have estimated  the BH mass  in LMC X-1 as
$m=(16\pm 1\times) [0.5/cos(i)]^{1/2}$ and they inferred that
$R_{\rm in}\approx 3R_S$. Using the mass $m=16$, the period $P=4.22$ d found 
by Hutchings et al. (1983) and the estimate of $x_{\rm in}=1$ and 
 $r_{\rm adj}=2$  (for high mass
accretion rate), we find from formula (14) that
$\nu_0=0.072$ Hz. This value of $\nu_0$  is very close to the narrow 
$0.08\pm 0.009$ Hz QPO peak found by Ebisawa et al. (1989)  with a harmonic
seen in the 1-16 keV range. 
The agreement of $\nu_0$ with the observed frequency may be viewed as a
confirmation of the correct mass estimate given 
by ShT99 for this source.

\centerline{\it GRO J1655-40}

Remillard et al. (1999) have found the relatively stable frequency near 
0.1 Hz which appears in the high-soft state.  
The mass $m\approx 7$  and 
the period $P=2.62$ d for this source  is known from the observations 
by Orosz \& Bailyn (1997).
Using these values in formula (14) and $x_{\rm in}=1$ (BOR99) and 
$r_{\rm adj}=2$ we find that $\nu_0=0.14$ Hz which is
comparable  with the observable value of $\nu_0=0.1$ Hz. 

\centerline{\it XTE J1118+480}

The source XTE J1118+480 is presumably a black hole candidate (Revnivtsev,
Sunyaev \& Borozdin 2000). 
The source is observed in the low hard state during the rising phase of the
2000 April outburst (Hynes et al. 2000). 
In this source, almost three identical QPO frequencies  
were detected in optical (0.1 Hz) , ultraviolet (0.08 Hz) 
by HST (Haswell et al. 2000) and in X-ray energy ranges
(0.08 Hz) by RXTE (Revnivtsev et al. 2000) and (0.11 Hz) 
 by ASCA (Yamaoka, Ueda \& Dotani 2000). 
The stability and independence of this frequency on the photon energy is 
a necessary feature of the GMD when the disk oscillates
as a whole body. We do not exclude that these oscillations of
outer part of disk may be amplified by the resonance effect due to reprocessing
of X-ray flux oscillating with the same frequency. In fact, a preliminary
 analysis by Haswell et al. (2000) indicates that ultraviolet variability 
lags are behind  X-ray by 1-2 s. This is likely due to the X-ray reprocessing
in the disk. 
The energy spectrum of  XTE J1118+480 is a typical thermal Comptonization
spectrum (Sunyaev \& Titarchuk 1980) with the best-fit parameters   
optical depth $\tau=3.2$ and $kT_e=33$ keV. 
%

Recently Wood et al. (2000) have demonstrated from extensive observations 
of XTE J1118+480 using the Unconventional Stellar Aspect (USA) experiment 
 and RXTE   that the QPO shows an upward drift from 0.07 Hz to 0.15 Hz while 
 the source intensity slowly rises and then decreases.
We can interpret this behaviour behavior of the QPO frequency by a significant change  
of the disk size $r_{out}$ during the spectral evolution of the source. 
The spectral evolution is related to the change of  the mass accretion rate 
in the disk (Chakrabarti \& Titarchuk 1995) which in turn causes 
the change of  the disk size. 
Soria (2000, private communication) argues that such a change is seen in 
GRO J1655-40. In other words, when the disk mass accretion rate drops 
the spectrum gets harder and the QPO frequency increases (see Eq. 8)
because the size of the disk ($r_{out}$) decreases. 
If our interpretation is correct we expect {\it that the QPO frequency
correlates  with the spectral hardness} but not with the
source flux.
   
Following  ShT99 we suggest that in the low-hard state the 
inner disk edge retreats approximately at 17 Shwarszchild radii due to
evaporation of the innermost part. 
From formula (16) we infer the BH mass  in XTE J1118+480  
to be 20 solar masses  with an accuracy 10\% taking  observed 0.1 Hz  as 
the GDM frequency,  $x_{\rm in}=17/3=5.7$ (ShT99), 
$P=4.1$ h (Uemura et al. 2000) and $r_{\it adj}=3$. 
This mass  is comparable that ShT99 have found in LMC X-1 and Nova Muscae.

\section{Comparison of the global mode oscillation frequency with 
the persistent low frequencies observed in a number of NS  sources}

\centerline{\it Her X-1}
Averintsev, Titarchuk \& Sheffer (1992) made a numerical simulation of the
process of occultation of the X-ray emission region on the neutron star
surface and compared their results with the observed X-ray emission of Her X-1.
They demonstrated  constraints for a number of geometrical parameters of the 
Hz Her/Her X-1 system; in particular, they found the position of the inner edge
of accretion disk at $R_{in}=(10\pm 2) R_{\rm NS}$. 
Taking  $R_{\rm in}=10~R_{\rm NS}$, $R_{\rm out}=1.69\times 10^{11}$ cm (HW83),
$P=1.7$ d (Deeter et al. 1991), $m=1.4$ and $r_{\rm adj}=3$  we find from 
(14) that $\nu_0\approx0.067$ Hz which is close to the QPO frequency $0.05$ Hz
 obtained by Boroson et al.(2000) by analyzing the power density spectrum of
HST UV data.

\centerline{\it 4U 1323-62, EXO 0748-676 and 4U 1746-37} 

The Low mass X-ray binaries (LMXBs) sources 4U 1323-62, 
EXO 0748-676 and 4U 1746-37 exhibit QPOs frequencies of about 1 Hz (see
references in Introduction). 
 It was noted by Jonker, van der Klis \& Wijnands (1999)
 that a medium modulating the radiation from a central source is  a 
promising explanation.  
Here we put forward the idea that the oscillations of X-ray radiation
with the low frequency 
can be a result of the GDM modulation of the X-ray radiation from 
a central source. The frequency of GDM is expected to be very stable because 
it mostly depends on the intrinsic characteristics of the system such as a mass 
and  a rotation period (or the disk size).
These characteristics  are comparable for the three aforementioned NS sources.
 The periods are
2.93,  3.82, 5.7 hours for 4U 1323-62, EXO 0748-676 and 4U 1746-37 
respectively (van Paradijs 1995). To estimate $\nu_0$ we take   $x_{\rm in}=1$
(TO99) and $m=1.4$ which is a typical value of the NS mass.
The dependence of $\nu_0$ on the period according to formula (14) is not strong,
$\nu_0\propto P^{-7/15}$, which means  almost the same value of $\nu_0$ 
for each of the sources within uncertainties caused by assumptions 
regarding the disk size and the adjustment radius. 
     
The stability of $\nu_0$ can exist until  the inner part of the disk is
disrupted  or just moved to the larger radii.
In principle magnetic force and radiation pressure can modify the restoring
force of the GDM. The related effects are amendable and they will be 
taken into account at later time.

\centerline{\bf 5. Conclusions}

In this {\it Letter} we have presented a model of the global disk oscillation
mode. Our analysis shows that the frequencies of this mode of  0.01-1 Hz
are closely related to the large ratio of sizes ($10^4$) of the disk and 
its inner portion. This ratio reduces the 
characteristic kHz Keplerian  frequency of the inner part of the disk 
by 3-4 orders of magnitude. 
{\it The GDM  is a vertical mode and thus it should be often seen from the 
systems with the high inclination angle}. However  due to X-ray reprocessing 
in the outer part of the disk and twisting of the disk
 there is also a chance of detecting these QPOs from systems with small
inclination angles.
 The GDM frequency is expected to be also {\it independent of 
the photon energy and should be seen  in all energy bands} that the 
disk emits. The particular part  of the viscous disk is presumably seen in the
specific  energy range  because the effective temperature of disk $T_{\rm eff}$
is a function of radius $R$, namely $T_{\rm eff}\propto R^{-3/4}$.
This {\it Letter}  offers a new method of BH mass determination 
based on the persistent low QPO frequency measurement; 
it supplements the X-ray energy spectrum method and the dynamical optical
method of  mass function determination.  This new approach in our view 
is very promising and it allows an independent verification of BH mass obtained
by other methods.

The authors acknowledge discussions with Chris Shrader and Bram Boroson and
thank Joe Fainberg for  valuable comments on earlier manuscript of this paper.

\clearpage

\begin{figure}
\caption{ Schematic picture illustrating the idea of the global disk
oscillation. The disk as a whole body oscillates under influence of the gravity
of the central source. 
\label{Fig.1}}
\end{figure}

\noindent
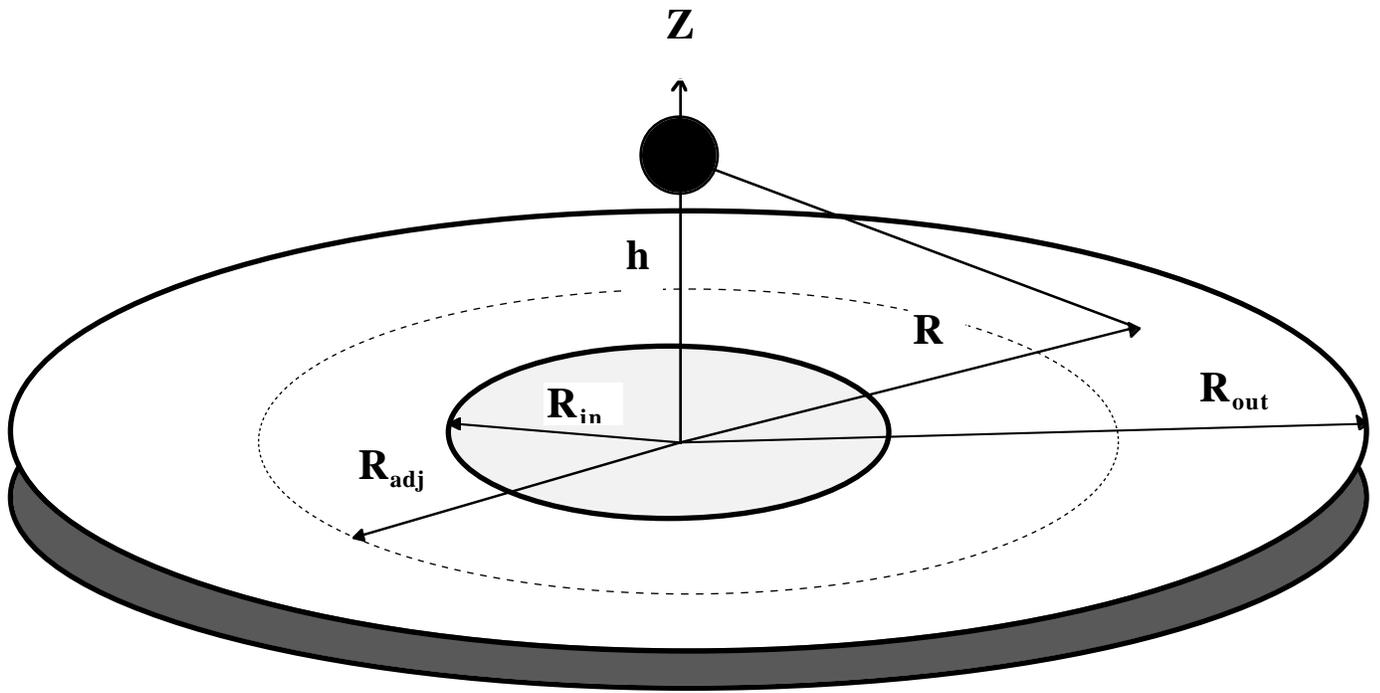
\end{document}